\documentclass[twocolumn,aps,prb,floats,showpacs,superscriptaddress,
preprintnumbers,amsmath,amssymb]{revtex4}

\usepackage{graphicx}
\usepackage{dcolumn}
\usepackage{bm}

\begin{document}


\title{Spin polarized current and
Andreev transmission in planar superconducting-ferromagnetic Nb-Ni junctions}

\author{E. M. Gonz\'alez}
\affiliation{Departamento de F{\'\i}sica de Materiales, Facultad de CC. F{\'\i}sicas, 
Universidad Complutense, 28040 Madrid, Spain} 
\author{A. Dom{\'\i}nguez-Folgueras}
\affiliation{Instituto de Ciencia de Materiales de Madrid, CSIC, Cantoblanco, 
28049 Madrid, Spain} 
\affiliation{Departamento de F\'{\i}sica, Universidad de Oviedo, 33007 Oviedo, Spain}
\author{F. J. Palomares} 
\affiliation{Instituto de Ciencia de Materiales de Madrid, CSIC, Cantoblanco, 
28049 Madrid, Spain} 
\author{R. Escudero}
\thanks{On leave from Instituto de Investigaci\'on en Materiales, Universidad Nacional Aut\'onoma de 
M\'exico, M\'exico D. F.}
\affiliation{Departamento de F{\'\i}sica de Materiales, Facultad de CC. F{\'\i}sicas, 
Universidad Complutense, 28040 Madrid, Spain} 
\author{J. E. Villegas}
\affiliation{Departamento de F{\'\i}sica de Materiales, Facultad de CC. F{\'\i}sicas, 
Universidad Complutense, 28040 Madrid, Spain} 
\author{J. M. Gonz\'alez}
\affiliation{Instituto de Ciencia de Materiales de Madrid, CSIC, Cantoblanco, 
28049 Madrid, Spain} 
\author{J. Ferrer}
\affiliation{Departamento de F\'{\i}sica, Universidad de Oviedo, 33007 Oviedo, Spain} 
\author{F. Guinea}
\affiliation{Instituto de Ciencia de Materiales de Madrid, CSIC, Cantoblanco, 
28049 Madrid, Spain} 
\author{J. L. Vicent}
\affiliation{Departamento de F{\'\i}sica de Materiales, Facultad de CC. F{\'\i}sicas, 
Universidad Complutense, 28040 Madrid, Spain}

\date{\today}

\begin{abstract}
We have measured and modeled, using three different approaches,
the tunneling current in a Nb/Nb$_x$O$_y$/Ni planar tunnel junction. The
experimental data could be fitted and the correct current
polarization could be extracted using a simple quasiclassical
model, even in the absence of an applied magnetic field. We also
discuss the microscopic structure of the barrier.
\end{abstract}


\maketitle

\section{Introduction}

Experiments with tunnel junctions using ferromagnetic metals\cite{Coleman}
have been an interesting topic since a long time. This subject has grown
again\cite{Akerman} because of the new field of spintronic where
spin dependent currents are an important requisite of many possible devices.
\cite{Prinz, Boeck} This implies that control and measurements of spin
polarized currents are needed. Spin-polarized electron tunneling\cite
{Meservey} is a key tool to measure the current polarization and to
understand the physics involved in these effects. Most of the recent
experimental works have been focused on the suppression of the Andreev
reflection and point contact geometry using a superconducting electrode.\cite
{Upadhyay, Soulen}However the local information extracted by point contact or
scanning tunneling microscope techniques seems to be less suitable
for devices than planar tunneling junctions. In addition, the
intrinsic difficulty of fabricating a perfect uniform oxide layer can
jeopardize the latter technique. Recently Kim and Moodera \cite{Kim} have
reported a large spin polarization of 0.25 from polycrystalline and
epitaxial Ni (111) films using Meservey and Tedrows's technique\cite{Meservey}
and standard Al electrode and oxide barriers.

In this work we show that the tunneling electron polarization of
ferromagnets could be extracted without applying a magnetic field to the
junction and using an oxide barrier with random metallic point contacts. The
experimental data are obtained for Nb/Nb$_x$O$_y$/Ni planar tunnel junctions. We
analyze possible models which describe the experimental results, with
emphasis on the information that can be obtained about the shape of the
barrier, and the relation between the polarization of the transmitted
current and the bulk polarization.

\section{Experimental method}

\begin{figure}[tbp]
\includegraphics{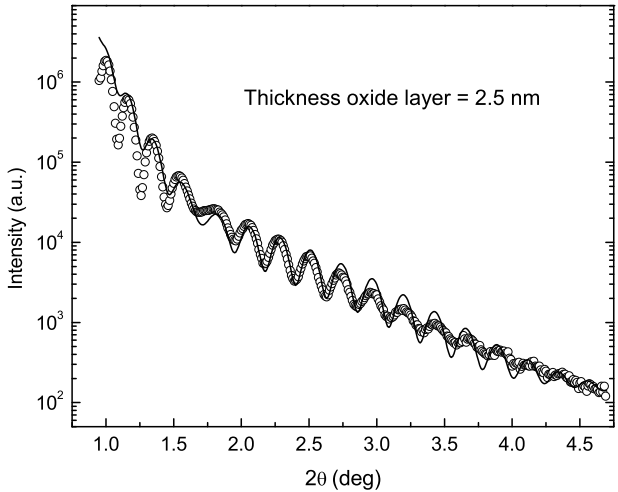}
\caption{\label{fig:xray1} 
Measured (open circles) and simulated (solid line) low-angle x-ray diffraction profiles of a
Nb film oxidized in a saturated water vapour atmosphere. The thickness of the
oxide layer is 2.5 nm, as extracted from the simulation.}
\end{figure}

\begin{figure}[tbp]
\includegraphics{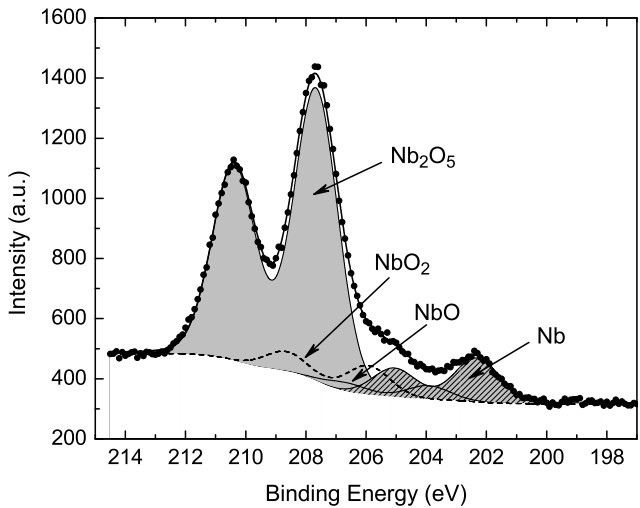}
\caption{\label{fig:xps1} 
X-ray photoelectron spectroscopy analysis of a Nb film oxidized 
in a saturated water vapour
atmosphere. Notice that the highest contribution component oxide
comes from Nb$_{2}$O$_{5}$.}
\end{figure}

Nb(110) and Ni(111) films, grown by dc magnetron sputtering,
were used as electrodes. The structural characterization of these films
was done by x-ray diffraction (XRD) and atomic force microscopy (AFM),
see for instance Villegas {\it et al}.\cite {PhysC369}
Briefly, the junction fabrication was as follows:
First, a Nb thin film of 100 nm thickness was evaporated on a Si substrate
at room temperature. An Ar pressure of 1 mTorr
was kept during the deposition. Under these conditions, the roughness of the
Nb film, extracted from XRD and AFM,
is less than 0.3 nm \cite{PhysC369} and superconducting critical
temperatures of 8.6 K are obtained. After this, the film was chemically
etched to make a strip of 1 mm width. A tunnel barrier was prepared by
oxidizing this Nb electrode in a saturated water vapour atmosphere at room
temperature.\cite{Halbritter} Fig. \ref{fig:xray1} shows the x-ray diffraction data (open
circles) of a Nb oxidized film. The thickness of the oxide layer was deduced
after the simulation of the data (solid lines) using the SUPREX program.\cite
{Fullerton} It can be seen that the oxide film thickness is 2.5 nm. 
X-ray photoelectron spectroscopy analysis performed in these oxidized films
reveals that dielectric Nb$_{2}$O$_{5}$ is the main oxide formed, as can
be seen in Fig. \ref{fig:xps1}. There are also other oxides, such as
metallic NbO, but in much less amount. Taken into account Grundner and
Halbritter studies, \cite{Halbritter} Nb$_{2}$O$_{5}$ is the outermost oxide
layer on Nb, whereas NbO is located closer to the Nb film. The
characterization by AFM reveals a RMS roughness of around 0.7 nm.

On top of this film (Nb with the oxide barrier), the second electrode of Ni
was deposited under the same conditions as Nb (up to 60 nm thickness) using
a mask to produce cross strips of 0.5 mm width, so that the overlap area $S$
of the two electrodes is 0.5 mm$^{2}$.

Perpendicular transport (tunneling configuration) was investigated by means
of characteristic dynamic resistance ($dV/dI$) versus voltage ($V$) using a
conventional bridge with the four-probe method and lock-in techniques. The
measured lock-in output voltage was calibrated in terms of resistance by
using a known standard resistor.

Experimental data (open circles) in Figs. 3, 4 and 6 show the behavior
of the normalized conductance $G(V)/G_n$ versus the bias voltage $V$. 
Conductance has been calculated as the inverse of the differential resistance
$dV/dI$ and the normalization has been done with respect to the dependent
voltage background conductance $G_n$. 

\section{Theoretical Description}

\subsection{Introduction}

Conductance $G$ across a normal-superconducting junction may be expressed as a
function of applied voltage $V$ in terms of the reflection coefficient $R$ of
electrons traversing the junction as
\begin{equation}
{{G(V)}\,=\,2\,\,{e}^2\,{v_F\,N_F\,S}\,\int\,{d\epsilon}\,\, (1+{A}-{R%
})\,\,\frac{df(\epsilon-eV)}{d\epsilon}}
\end{equation}
where ${f(\epsilon)}$ is the Fermi function; ${v_F}$ and ${N_F}$,
the Fermi velocity and density of states at the Fermi level,
respectively; and $A$ is the coefficient for Andreev reflection processes
whereby an electron with energy smaller than the gap impinging onto the
junction picks an electron of opposite spin to form a Cooper pair inside the
superconductor, leaving behind a hole. Andreev reflection processes are
proportional to the square of the conventional transmission coefficient of
the barrier $T$, and therefore are strongly supressed for highly resistive
barriers. Junctions with transmission coefficients smaller than about 0.1
show small subgap conductances.

Our experimental results, shown in Fig. \ref{fig:ModI}, exhibit a significant conductance
below the superconducting gap even at the lowest temperatures. Therefore we
expect that our effective oxide barriers should be neither too high nor too
thick.

We have used three models to describe the transmission across the oxide
barrier: i) The simple generalization of the Blonder-Tinkham-Klapwijk (BTK)
model\cite{BTK} to ferromagnetic electrodes proposed by Strijkers and
coworkers;\cite{Ji} ii) a generalization of the BTK model to
ferromagnetic electrodes with finite bulk magnetization; and iii) a
description of the effects of a finite current polarization in terms of spin
dependent transmission coefficients, in a similar way as discussed by 
P\'erez-Willard {\it et al}.\cite{Perez-Willard}.

\subsection{Strijkers' model (Model I)}

Strijkers' model uses as adjustable parameters the current polarization
\begin{equation}
{{P_{c}}=\frac{G_{\uparrow }(0)-G_{\downarrow }(0)}{G_{\uparrow
}(0)+G_{\downarrow }(0)},}
\end{equation}
the height of the barrier $Z$, which is modeled by a delta function, and the
size of the superconducting gap at the interface $\Delta $. The process of
electron transfer across the junction is split into a fully polarized
channel, for which the Andreev reflection is zero, and a paramagnetic
channel described by the BTK model in its usual form. The total conductance
is then written in terms of the conductance of the fully polarized channel (${G_{P}}$)
and the conductance of the paramagnetic channel (${G_{N}}$):
\begin{equation}
{{G\left( V\right) =\left( 1-P_{c}\right) G_{N}\left( V\right)
+P_{c}\,G_{P}\left( V\right) },}  \label{eq_conductance}
\end{equation}
This model interpolates between the paramagnetic case (BTK model), and the
half metal, where it predicts correctly that the amplitude for Andreev
reflection vanishes.

\begin{figure}[tbp]
\includegraphics{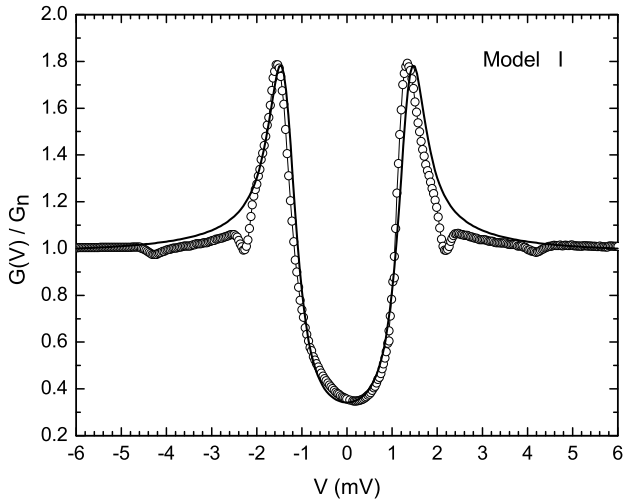}
\caption{\label{fig:ModI} 
Normalized tunneling conductance as a function of voltage of a Nb-Ni planar junction. 
Symbols correspond to experimental data and continuos solid line to the calculated curve using
Model I. The temperature of the experiment was $1.52$ K and the
fitting parameters were $Z=1.15$, $P_c=0.03$, $\Delta=1.4$ meV (see text for details).}
\end{figure}

The best fit obtained with this model is shown in Fig. \ref{fig:ModI}.
The parameters used are $\Delta = 1.4$ meV (superconducting gap), ${{Z} = 1.15}$, 
and ${{P_c} = 0.03}$. This value of the polarization,
however, is well below the usually reported current polarization of Ni.

\subsection{Generalization of BTK model for a ferromagnetic electrode (Model II)}

We next introduce ferromagnetism through an exchange splitting $J$ in one of
the electrodes. Therefore wave-vectors depend on spin as ${{\hbar
\,k_{\sigma }\,=\,(2\,m\,(E_{F}\,+\,\sigma \,J/2))^{1/2}}}$. Transmission
across the barrier is therefore spin-dependent, and Andreev reflection mixes
both spins, giving different normal and Andreev reflection probabilities for
each spin. Conductance can be calculated for each spin in the same way as in
the BTK model using these reflection probabilities. Total conductance will
be then $G=G_{\uparrow }+G_{\downarrow }$, where
\begin{equation}
{{G_{\sigma }}=2\,\,{e}^{2}\,{v_{F\sigma }\,N_{F\sigma }\,S}\,\int \,{%
d\epsilon }\,\,(1+\frac{k_{-\sigma }}{k_{\sigma }}\,{A}_{\sigma }-{R}%
_{\sigma })\,\,\frac{df(\epsilon -eV)}{d\epsilon }}  \label{eq_conductance2}
\end{equation}
The adjustable parameters in this model are $J$, the height of the barrier for
zero splitting $Z$, again modeled by a delta function, and the gap at the
interface $\Delta$.

\begin{figure}[tbp]
\includegraphics{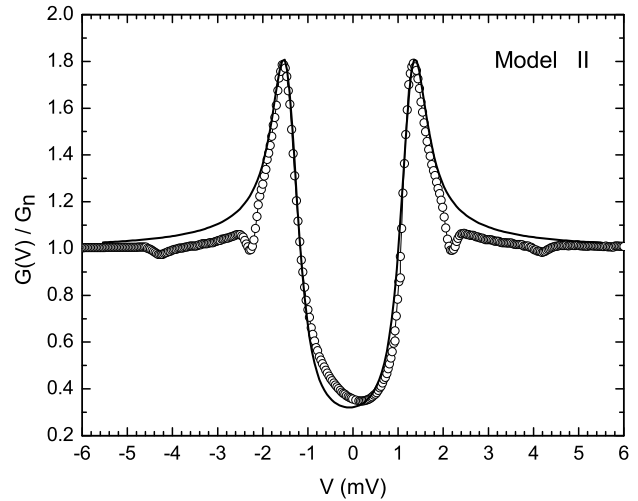}
\caption{\label{fig:ModII} 
Fit to the experimental results obtained with model II (see text for details).
The fitting parameters were $Z=1.13$, $\Delta=1.4$ meV, $J=0.8$ eV and 
the transmission coefficients were
$T_\uparrow = 0.48$ and $T_\downarrow = 0.37$.}
\end{figure}

This model leads to the results shown in Fig. \ref{fig:ModII}. A good fit is
obtained using ${{Z} = 1.13}$, $\Delta = 1.4$ meV and $J = 0.8$ eV,
which agrees well with the electronic band structure of Ni. This
exchange provides a bulk polarization
\begin{equation}
{{P_b = \frac{n_\uparrow - n_\downarrow}{n_\uparrow + n_\downarrow}
\approx \frac{J}{2\,E_F}} \approx 0.4}  \label{eq_bulkpolarization}
\end{equation}
if we take ${{E_F}\simeq 1}$ eV, which is the botton of the band of
Ni along our experimental $\Gamma \mathrm{L}$ direction. The model also
allows us to define the current polarization, assuming a unidimensional
barrier, in terms of the difference in the transmission coefficients through
the barrier, ${{T}_\uparrow = 0.48}$ and ${{T}_\downarrow =
0.37}$. We find
\begin{equation}
{{P_c \approx \frac{T_\uparrow - T_\downarrow}{T_\uparrow +
T_\downarrow}} \approx 0.13 ,}  \label{eq_polarization}
\end{equation}
which we still feel to be somewhat too low.

\begin{figure}[tbp]
\includegraphics{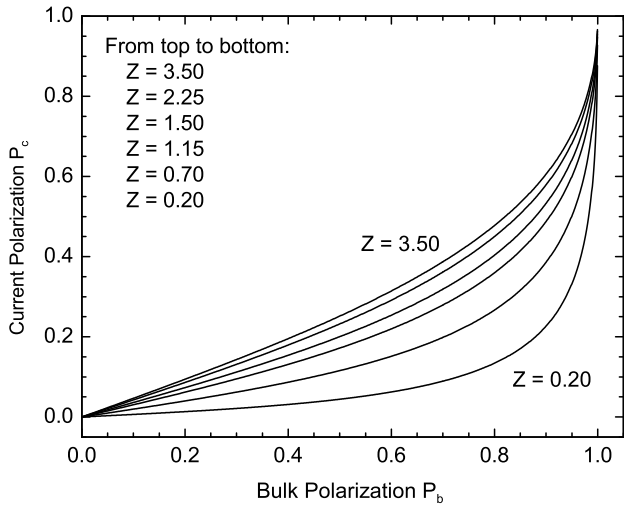}
\caption{\label{fig:PcPb1} 
Polarization determined from the transmission coefficients across the barrier,
eq.(\protect{\ref{eq_polarization}}) as a function of the bulk polarization of the
ferromagnetic electrode.}
\end{figure}

We note that the polarization as determined from the transmission
coefficients needs not agree with the bulk polarization of the
ferromagnetic electrode.\cite{MacLaren, Mazin} Similar effects can arise
when the conduction
bands are built up of localized and delocalized atomic orbitals.\cite{Teresa}
A comparison between the polarization defined using Eq.(\ref
{eq_polarization}) and the bulk polarization, for barrier strenghts in the
range from 0.2 to 3.5 is shown in Fig. \ref{fig:PcPb1}. In the limits ${{P_b} \rightarrow 0}$
and ${{P_b} \rightarrow 1}$ both
polarizations are equal, but for intermediate values current polarization is
always smaller than bulk one.

\subsection{Simple quasiclassical theory (Model III)}

We now use quasiclassical theory, with boundary conditions dumped into the
transmission coefficients ${{T}_\sigma}$. Conductance
is then determined using Eq. (\ref{eq_conductance2}) with ${k_\uparrow=k_\downarrow}$,
where the effective reflection coefficients are
\begin{eqnarray}
{ A_\sigma}&=&{T_\sigma\,\,T_{-\sigma}\,\,\left|
\frac{f}{1+r_\sigma\,r_{-\sigma}+(1-r_\sigma\,r_{-\sigma})\,\,g}\right|^2}
\nonumber\\
{R_\sigma}&=&{\left|\frac{r_\sigma+r_{-\sigma}+(r_\sigma-r_{-\sigma})\,\,g}
{1+r_\sigma\,r_{-\sigma}+(1-r_\sigma\,r_{-\sigma})\,\,g}\right|^2}.
\end{eqnarray}

Here ${g}$ and ${f}$ are the (spin independent) normal and
anomalous components of the Green's functions evaluated right at the
interface, at the superconducting side, and ${{r_\sigma^2+T_\sigma}=1}$.

This scheme is, in principle, exact, although it does not allow us to
extract information about the nature of the barrier, which is treated as a
black box whose internal structure is ignored. We also assume that the
electrodes are in the clean limit, to simplify the calculation. Once the
transmission coefficients across the barrier which best fit the experimental
data have been determined, we have performed calculations in the normal
regime with more realistic square barriers in order to check the physical
properties of the model. Neglecting effects related to the dispersion in the
direction parallel to the junction, the results obtained using this model
should be reproduced by the generalized BTK model with the appropiate choice
of transmission coefficients.

\begin{figure}[tbp]
\includegraphics{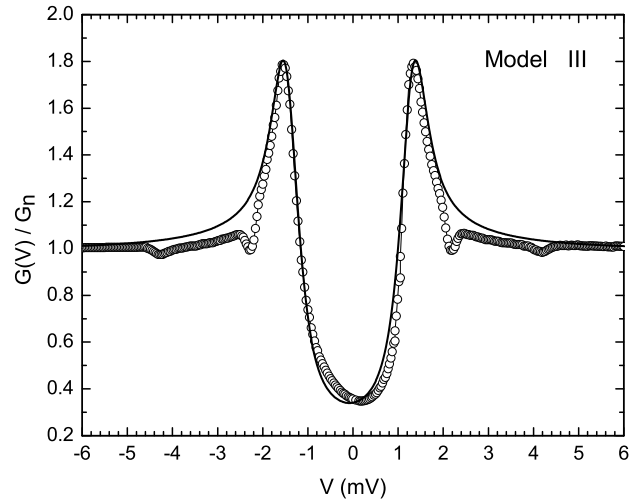}
\caption{\label{fig:ModIII} 
Fit to the experimental results obtained with model III (see text for details). 
The fitting parameters were $\Delta=1.38$ meV and the used
transmission coefficients were $T_\uparrow = 0.5$ and $T_\downarrow = 0.32$.}
\end{figure}

The best results obtained using this model are shown in Fig. \ref{fig:ModIII}.
The parameters used are $\Delta = 1.38$ meV, ${{T}_\uparrow = 0.5}$
and ${{T}_\downarrow = 0.32}$. Using the expression in eq.(\ref
{eq_polarization}) we obtain a current polarization ${{P_c} = 0.22 }$.

\subsection{Characterization of the barrier}

\begin{figure}[tbp]
\includegraphics{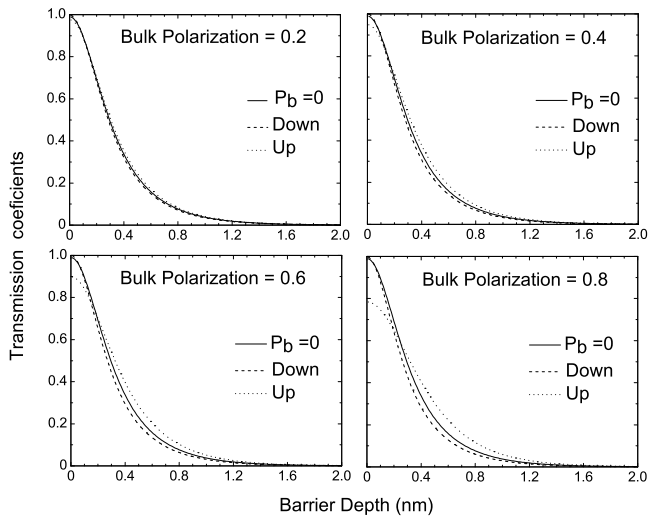}
\caption{\label{fig:trans1} 
Transmission coefficients as a function of barrier depth for a barrier of heigth $U=0.1$ eV.
Bulk polarization in the ferromagnetic electrode is, from left to right and from top to
bottom: 0.2, 0.4, 0.6, 0.8, as indicated in each graph.}
\end{figure}

We have also related the transmission coefficients used in the fits with the
thickness $d$ and height $U$ of more realistic square barriers. Grundner and
Halbritter have found that the effective height of the barrier of Nb$_{2}$O$%
_{5}$ oxide layers is of order ${{U}\approx 0.1}$ eV.\cite{Halbritter}
Fig. \ref{fig:trans1} shows the transmission through a barrier of
such height as function of d, for different values of bulk polarization. The
Fermi energy in the Ni ferromagnetic electrode is approximately ${{E_{F}}\approx 1}$ eV,
while in the Nb superconducting electrode is of about ${{E_{F}}\approx 1.5}$ eV
along the $\Gamma \mathrm{P}$ direction. We
find transmission coefficients comparable to those required to fit the
experimental data for barrier depths d $\sim 5$ \AA .

The significant conductance observed at voltages below the gap implies the
existence of good contacts. One may get an estimate of the number of these
contacts by dividing the quantum unit of resistance, ${{h/(2\,e^{2})}}$
by the observed resistance, ${{R}=20\,\Omega }$ (the fact that the
transmission coefficients needed to fit the data are below 1 does not alter
the order of magnitude). Assuming that the area of each of these conducting
channels is of the order of a few tens of \AA $^{2}$ we obtain that the area
of these contacts is a small fraction of the area of the oxide barrier ($10^{-8}$),
suggesting that the conductance is mainly due to tiny spots where
the barrier is narrower than the average thickness of the oxide layer.
The existence of zones where the barrier is $\sim 5$ \AA \thinspace\ thick
is not inconsistent with the experimentally observed corrugation, and agrees
with our estimation of the barrier depth.

\section{Summary and conclusions}

The experimental results can be fit with different models. The simplest one,
which describes the junction in terms of a paramagnetic and a fully
polarized channel, requires as an input a bulk polarization which is much
lower than the observed polarization of Ni. A more accurate one is the exact
generalization of the BTK model, which describes properly the bulk
polarization of Ni but not its current one. A scattering approach, which
does not require knowledge of the structure of the barrier, describes well
the experimental data with reasonable values for the transmission
coefficients for majority and minority spins, giving the correct value of
the current polarization. The experimental data show an additional
contribution which is not symmetric with respect to the applied voltage,
this small asymmetry is not relevant for our results. This contribution may
arise from inelastic scattering of magnons at the interface.\cite{Tkchakov}

The polarization inferred from the transmission coefficients calculated with
the BTK model does not coincide with the bulk magnetization of the
ferromagnetic electrode, used as an input. We have studied the relation
between these two quantities, showing a significant dependence on the height
of the barrier.

\begin{acknowledgments}
This work was supported by C.I.C.Y.T. (MAT2002-04543,
MAT 2002-02219, MAT2002-0495-C02-01, BFM2003-03156) and Ram\'on Areces Foundation.
E.M.G. acknowledges Ministerio de Ciencia y Tecnolog{\'\i}a
for a Ram\'on y Cajal contract. A.D.-F. thanks Ministerio 
de Educaci\'on, Cultura y Deporte
for a FPU grant (AP2002-1383). 
R. Escudero thanks Universidad Complutense and Ministerio
de Educaci\'on, Cultura y Deporte for a sabbatical professorship. 
\end{acknowledgments}


\end{document}